\documentclass[epj,nopacs]{svjour}
\usepackage{amssymb, amsmath}
\usepackage{mathbbold}
\usepackage{graphicx}% Include figure files
\usepackage{dcolumn}% Align table columns on decimal point
\usepackage{bm}% bold math
\usepackage{hyperref}% add hypertext capabilities
\usepackage{color}
\begin{document}

\title{Meson Screening Mass in a Strongly Coupled Pion Superfluid}
\author{Yin Jiang, Ke Ren, Tao Xia \and Pengfei Zhuang}
\institute{Physics Department, Tsinghua University, Beijing
100084, China}
\date{\today}
\abstract{
We calculate the meson screening mass in a pion superfluid in the
framework of Nambu--Jona-Lasinio model. The minimum of the
attractive quark potential is always located at the phase boundary
of pion superfluid. Different from the temperature and baryon
density effect, the potential at finite isospin density can not be
efficiently suppressed and the matter is always in a strongly
coupled phase due to the Goldstone mode in the pion superfluid.}
\maketitle
%\pacs{}
%\linenumbers
%\tableofcontents

%%%%%%%%%%%%%%%%%%%%%%%%%%%%%%%%%%%%%%%%%%%%%%%%%%%%%%%%%%%%%%%%%%%%%%%
\section{Introduction}
Understanding the behavior of quantum chromodynamics (QCD) at finite
temperature and density is essential for a description of the
development of the early universe and compact stars and for the
explanation of the results from high energy nuclear collisions at
Relativistic Heavy Ion Collider (RHIC) and Large Hadron Collider
(LHC), where high temperature and/or density can be reached. From
Lattice simulations~\cite{karsch1,karsch2,cheng,borsanyi} at finite
temperature, there are two QCD phase transitions, namely the
deconfinement transition from hadronic matter to quark matter and
the chiral transition from chiral symmetry breaking to its
restoration.

At finite baryon density, the QCD phase structure becomes more rich.
Besides the deconfinement and chiral phase transitions, the color
symmetry which is strict in vacuum is spontaneously broken and leads
to a new phase at high density, the so-called color
superconductor~\cite{barrois,bailin1,bailin2} which may exist in the
core of dense quark stars. However, due to the fermion sign
problem~\cite{muroya}, there is not yet precise lattice result at
finite baryon density. While the properties of the new phases at
extremely high density can be studied with perturbative QCD, to
investigate the phase transitions themselves at moderate baryon
density one needs effective QCD models at low energy.

The isospin symmetry is spontaneously broken when the isospin
chemical potential is larger than the pion mass, resulting in the
so-called pion superfluid phase~\cite{migdal} which may exist in the
core of neutron stars. In comparison with the temperature and baryon
density effect, the study on isospin effect or pion superfluid phase
includes new phenomena: 1) The isospin structure of QCD phase
transitions can be directly examined through lattice simulations
without serious technical problems~\cite{kogut1,kogut2} emerged at
finite baryon density; 2) While for the color superconductivity at
finite baryon density the Goldstone modes corresponding to the local
color symmetry breaking are eaten up by the Higgs mechanism, the
Goldstone mode corresponding to the global isospin symmetry breaking
dominates the thermal and dynamic properties of the pion
superfluid~\cite{mu1}; 3) While the phase transitions of
deconfinement and chiral symmetry restoration occur at finite
temperature and baryon density, there might be no
deconfinement~\cite{son} and the chiral structure is also
significantly changed~\cite{he} at finite isospin density.

From the lattice calculated thermodynamics at finite temperature,
the energy density~\cite{cheng} and pressure~\cite{borsanyi} do not
reach the corresponding ideal gas limit in the deconfined phase,
this means that the quark matter close to the phase transition is
not a weakly but a strongly interacting system. The strongly coupled
quark matter predicted by the theoretical calculations is recently
supported by the strong collective flow~\cite{star1,star2,phenix1}
observed in heavy ion collisions at RHIC. It is widely accepted that
the coupling strength will decrease with increasing temperature and
baryon density, and the quark matter will be in a weakly coupled
phase when the temperature or density is high enough.

The potential between two quarks in hot and dense medium can well
describe the changes in quark properties at finite temperature and
density. It presents a direct way to understand whether the matter
is a strongly coupled one. For instance, the heavy quark potential
extracted from lattice simulations~\cite{kaczmarek} at finite
temperature is the key point to explain the $J/\psi$
suppression~\cite{phenix2} observed in heavy ion collisions at RHIC.
At finite temperature and baryon density, the effect of chiral
symmetry restoration on the quark potential is
investigated~\cite{mu2} in the framework of Nambu--Jona-Lasinio
(NJL) model~\cite{njl}. It is found that the minimum of the
attractive potential is located at the critical temperature $T_c$ or
the baryon chemical potential $\mu_B^c$ of the chiral phase
transition, and the strongly coupled matter survives only in the
temperature region $1<T/T_c\lesssim 2-3$ or baryon density region
$1<\mu_B/\mu_B^c\lesssim 2$. In this paper, we study the meson
screening mass and the quark potential at finite isospin density in
the NJL model. We will focus on the potential in the pion superfluid
and explain its surprising isospin behavior induced by the Goldstone
mode corresponding to the spontaneous isospin symmetry breaking.

The original NJL model~\cite{njl} is inspired by the BCS theory
describing normal electron superconductor, and therefore its version
at quark level is widely used to describe chiral symmetry
restoration, color superconductor and pion superfluid (For reviews
and general references, see
Ref.~\cite{vogl,klevansky,volkov,hatsuda,buballa,ebert1,ebert2}). To
understand the results of lattice QCD thermodynamics in terms of
quasi-particle degrees of freedom, the model is recently extended to
include Polyakov loop dynamics
(PNJL)~\cite{meisinger,fukushima,ratti}. Because of the contact
interaction among quarks, there is no confinement in the model and
it is necessary to introduce a momentum cutoff $\Lambda$ to avoid
the singularity in momentum integrations. From the uncertainty
principle, the minimal length scale in the NJL model is $R\sim
1/\Lambda$. Taking the standard cutoff value $\Lambda \sim 600$
MeV~\cite{vogl,klevansky,volkov,hatsuda,buballa}, we have $R \sim
1/3$ fm. This means that, the quark potential calculated in the NJL
model might be reasonable in the region of $r>1/3$ fm. In the short
range the model is probably not applicable.

The paper is organized as follows. In Section \ref{s2} we review the
meson propagators in the pion superfluid in the NJL model and
calculate the meson screening masses. In Section \ref{s3} we
calculate the quark potential exactly and in pole approximation,
analyze the dominant contribution from the Goldstone mode due to
spontaneous isospin symmetry breaking, and discuss the strongly
coupled matter in the whole pion superfluid. We summarize in Section
\ref{s4}.

%%%%%%%%%%%%%%%%%%%%%%%%%%%%%%%%%%%%%%%%%%%%%%%%%%%%%%%%%%%%%%
\section{meson screening masses}
\label{s2}

To simplify the numerical calculations, we take the original NJL
model and neglect the Polyakov loop
potential~\cite{meisinger,fukushima,ratti} which disappears at zero
temperature. The two flavor NJL model is defined through the
Lagrangian density
\begin{equation}
\label{njl1} {\mathcal{L}} = \overline\psi\left(i\gamma^\mu
\partial_\mu-m_0+\mu\gamma_0\right)\psi+G\left[\left(\overline\psi\psi\right)^2+\left(\overline\psi
i\gamma_5{\bf {\tau}}\psi\right)^2\right]
\end{equation}
with scalar and pseudoscalar interactions corresponding to $\sigma$
and ${\bf \pi}$ excitations, where $m_0$ is the current quark mass,
$G$ is the coupling constant with dimension (GeV)$^{-2}$, the quark
fields $\psi$ and $\overline\psi$, the Pauli operators ${\bf \tau} =
(\tau_1,\tau_2,\tau_3)$ and the quark chemical potential
$\mu=diag(\mu_u,\ \mu_d)=diag(\mu_B/3+\mu_I/2,\mu_B/3-\mu_I/2)$ are
matrices defined in flavor space, and $\mu_B$ and $\mu_I$ are baryon
and isospin chemical potentials. At $\mu_I=0$, the system has the
symmetry $U_B(1)\bigotimes SU_I(2)\bigotimes SU_A(2)$, corresponding
to baryon number symmetry, isospin symmetry and chiral symmetry.
However, the isospin symmetry $SU_I(2)$ is explicitly broken down to
$U_I(1)$ global symmetry at small $\mu_I$, and then the $U_I(1)$
symmetry is further spontaneously broken down with condensation of
charged pions at large $\mu_I$. At $\mu_B=0$, the Fermi surfaces of
$u (d)$ and anti-$d (u)$ quarks coincide and hence the condensate of
$u$ and anti-$d$ quarks is favored at $\mu_I>0$ and the condensate
of $d$ and anti-$u$ quarks is favored at $\mu_I <0$. A finite
$\mu_B$ provides a mismatch between the two Fermi surfaces and will
reduce the pion condensation.

Introducing the chiral condensate
\begin{equation}
\label{chiral} \sigma=\langle\bar\psi\psi\rangle
\end{equation}
and pion condensate
\begin{equation}
\label{pion} \pi=\sqrt 2\langle\bar\psi i\gamma_5\tau_+\psi\rangle =
\sqrt 2\langle\bar\psi i\gamma_5\tau_-\psi\rangle
\end{equation}
with $\tau_{\pm} = \left(\tau_1 \pm i\tau_2\right)/\sqrt{2}$, which
are respectively the order parameters for the chiral phase
transition and pion superfluid, the inverse of the quark propagator
in mean field approximation defined in flavor space
\begin{equation}
\label{quark}
{\cal S}(p)= \left(\begin{array}{cc} {\cal S}_{uu}(p)&{\cal S}_{ud}(p)\\
{\cal S}_{du}(p)&{\cal S}_{dd}(p)\end{array}\right)
\end{equation}
can be written as
\begin{equation}
{\cal S}^{-1}(p)=\left(\begin{array}{cc} \gamma^\mu p_\mu+\mu_u\gamma_0-m & 2iG\pi\gamma_5\\
2iG\pi\gamma_5 & \gamma^\mu p_\mu+\mu_d\gamma_0-m\end{array}\right),
\end{equation}
where $m=m_0-2G\sigma$ is the dynamic quark mass induced by the
spontaneous chiral symmetry breaking. With the quark propagator, the
physical condensates $\sigma$ and $\pi$ are determined by the gap
equations,
\begin{eqnarray}
\label{gaps} \sigma &=& -i\int{d^4p\over
(2\pi)^4}\text{Tr}\left[{\cal S}_{uu}(p)+{\cal
S}_{dd}(p)\right],\nonumber\\
\pi &=& \int{d^4p\over (2\pi)^4}\text{Tr}\left[\left({\cal
S}_{ud}(p)+{\cal S}_{du}(p)\right)\gamma_5\right]
\end{eqnarray}
with the trace taken in color and Dirac spaces, where the four
dimensional momentum integration is defined as $\int d^4p/(2\pi)^4=iT\int
d^3{\bf p}/(2\pi)^3\sum_n$ in the imaginary time formalism of finite
temperature field theory with the Matsubara frequency
$\omega_n=(2n+1)\pi n T,\ n=0,\pm 1,\pm 2,...$ for fermions.

In the NJL model, the meson modes are regarded as quantum
fluctuations above the mean field and can be effectively expressed
at quark level in terms of quark bubble summation in random phase
approximation (RPA)~\cite{vogl,klevansky,volkov,hatsuda,buballa}. In
normal phase without pion condensation, the bubble summation selects
its specific isospin channel by choosing at each stage the same
proper polarization. In the pion superfluid phase, however, the
quark propagator contains off-diagonal elements in flavor space, we
must consider all possible isospin channels in the bubble summation.
In this case, the meson polarization function becomes a matrix in
the four dimensional meson space with off diagonal
elements~\cite{he}
\begin{equation}
\label{bubble} \Pi_{jk}(q) = i\int{d^4 p\over (2\pi)^4} \text{Tr}
\left[\Gamma_j^* {\cal S}\left(p+{q\over 2}\right)\Gamma_k {\cal
S}\left(p-{q\over 2}\right)\right]
\end{equation}
with $j,k = \sigma,\ \pi_+,\ \pi_-,\ \pi_0$, where the trace is
taken in color, flavor and Dirac spaces, and the meson vertexes are
defined as
\begin{equation}
\label{vertex} \Gamma_j = \left\{\begin{array}{ll}
1 & j=\sigma\\
i\tau_+\gamma_5 & j=\pi_+ \\
i\tau_-\gamma_5 & j=\pi_- \\
i\tau_3\gamma_5 & j=\pi_0\ ,
\end{array}\right.\ \ \
\Gamma_j^* = \left\{\begin{array}{ll}
1 & j=\sigma\\
i\tau_-\gamma_5 & j=\pi_+ \\
i\tau_+\gamma_5 & j=\pi_- \\
i\tau_3\gamma_5 & j=\pi_0. \\
\end{array}\right.
\end{equation}

From the definition (\ref{bubble}), the polarization function $\Pi$
is a symmetric matrix in the meson space with $\Pi_{jk}=\Pi_{kj} $,
and the neutral pion $\pi_0$ is decoupled from the charged pions
$\pi_+$ and $\pi_-$ and the isospin singlet $\sigma$ with $
\Pi_{\pi_0\sigma}=\Pi_{\pi_0\pi_+}=\Pi_{\pi_0\pi_-}=0$. Considering
further the relations $\Pi_{\sigma\pi_-}(q)=\Pi_{\sigma\pi_+}(-q)$
and $\Pi_{\pi_-\pi_-}(q)=\Pi_{\pi_+\pi_+}(-q)$, there are only five
independent polarization elements $\Pi_{\sigma\sigma}$,
$\Pi_{\sigma\pi_+}$, $\Pi_{\pi_+\pi_+}$, $\Pi_{\pi_+\pi_-}$ and
$\Pi_{\pi_0\pi_0}$. Taking the trace and performing the fermion
frequency summation in (\ref{bubble}), and introducing the following
definitions based on the quark energy $E_p = \sqrt{{\bf p}^2+m^2}$,
$\epsilon^{ab}_p =a\sqrt{\left(E_p+b\mu_I/2\right)^2+4G^2\pi^2}$,\\
$u^{ab}_p =\sqrt{\left(1+ \left(E^b_p+\mu_I/2\right)/
\epsilon^{ab}_p\right)/2}$, $E^a_p = a E_p$, $v_p^{ab}=a u_p^{ab}$ with $a,b=\pm$ and
the Fermi-Dirac distribution function $f(x) = 1/\left(
e^{(x-\mu_B/3)/T}+1\right)$, the complicated expressions for the
polarizations are greatly simplified, and the five independent
elements can be explicitly expressed in a compact way in (9),
\newcounter{mytempeqncnt}
\begin{figure*}[!t]
\normalsize
\setcounter{equation}{8}
\begin{eqnarray}
\label{polarization}
&& \Pi_{\sigma\sigma}(q) = -3\int{d^3{\bf
p}\over (2\pi)^3} \sum_{a,b,c,d=\pm} \left(1+{m^2-{\bf p}_+\cdot
{\bf p}_-\over E_{p_+}^a E_{p_-}^b}\right)\left(v^{ca}_{p_+}u^{db}_{p_-} +
u^{(-c)a}_{p_+}v^{(-d)b}_{p_-}\right)^2
{f(\epsilon^{(-c)a}_{p_+})-f(\epsilon^{(-d)b}_{p_-})\over q_0+\epsilon^{db}_{p_-}-\epsilon^{ca}_{p_+}},\nonumber\\
&& \Pi_{\sigma\pi_+}(q) = -3\sqrt 2\int{d^3{\bf p}\over (2\pi)^3}
\sum_{a,b,c,d=\pm} \left({m\over E_{p_+}^a}+{m\over
E_{p_-}^b}\right)u^{ca}_{p_+}u^{(-d)b}_{p_-}\left(v^{(-c)a}_{p_+}u^{(-d)b}_{p_-} +
u^{ca}_{p_+}v^{db}_{p_-}\right)
{f(\epsilon^{(-c)a}_{p_+})-f(\epsilon^{(-d)b}_{p_-})\over q_0+\epsilon^{db}_{p_-}-\epsilon^{ca}_{p_+}},\nonumber\\
&& \Pi_{\pi_+\pi_+}(q) = 6\int{d^3{\bf p}\over (2\pi)^3}
\sum_{a,b,c,d=\pm} \left(1+{m^2+{\bf p}_+\cdot {\bf p}_-\over
E_{p_+}^a E_{p_-}^b}\right)
\left(u^{ca}_{p_+}u^{(-d)b}_{p_-}\right)^2
{f(\epsilon^{(-c)a}_{p_+})-f(\epsilon^{(-d)b}_{p_-})\over q_0+\epsilon^{db}_{p_-}-\epsilon^{ca}_{p_+}},\nonumber\\
&& \Pi_{\pi_+\pi_-}(q) = -6\int{d^3{\bf p}\over (2\pi)^3}
\sum_{a,b,c,d=\pm} \left(1+{m^2+{\bf p}_+\cdot {\bf p}_-\over
E_{p_+}^a E_{p_-}^b}\right)
v^{ca}_{p_+}u^{(-c)a}_{p_+}v^{db}_{p_-}u^{(-d)b}_{p_-}
{f(\epsilon^{(-c)a}_{p_+})-f(\epsilon^{(-d)b}_{p_-})\over q_0+\epsilon^{db}_{p_-}-\epsilon^{ca}_{p_+}},\nonumber\\
&& \Pi_{\pi_0\pi_0}(q) = -3\int{d^3{\bf p}\over (2\pi)^3}
\sum_{a,b,c,d} \left(1-{m^2+{\bf p}_+\cdot {\bf p}_-\over E_{p_+}^a
E_{p_-}^b}\right) \left(v^{ca}_{p_+}u^{db}_{p_-} +
u^{(-c)a}_{p_+}v^{(-d)b}_{p_-}\right)^2
{f(\epsilon^{(-c)a}_{p_+})-f(\epsilon^{(-d)b}_{p_-})\over
q_0+\epsilon^{db}_{p_-}-\epsilon^{ca}_{p_+}}
\end{eqnarray}
\setcounter{equation}{9}
\hrulefill
\vspace*{4pt}
\end{figure*}
with ${\bf p}_\pm={\bf p}\pm{{\bf q}/2}$. Note that there is no
Lorentz invariance at finite temperature and density, and the
polarization elements are no longer functions of the Lorentz scalar
quantity $q^2=q_0^2-{\bf q}^2$ but depend separately on $q_0^2$ and
${\bf q}^2$.

Describing the meson exchange in quark scattering via the quark
bubble summation and taking into account the spontaneous
isospin symmetry breaking, the meson propagator can effectively be
expressed as a matrix in the four dimensional meson space~\cite{he}
\begin{equation}
\label{meson} D\left(q_0^2,{\bf
q}^2\right)=\frac{-2G}{1-2G\Pi\left(q_0^2,{\bf q}^2\right)}.
\end{equation}
The dynamical meson mass $M_j$ in vacuum is defined as the pole of
the meson propagator at $q^2=M_j^2$. At finite temperature and
density, the position of the pole is changed to $q_0^2=M_j^2, {\bf
q}^2=0$,
\begin{equation}
\label{pole1} \text{det}\left[1-2G\Pi\left(M_j^2,0\right)\right]=0,
\end{equation}
where the symbol det means the determinant of the matrix $1-2G\Pi$.
In normal phase without pion condensation, it is simplified to four
independent mass equations, $1-2G\Pi_{jj}\left(M_j^2, 0\right)=0$.
At $\mu_I=0$ and in chiral limit, it is found~\cite{zhuang1} that
the three pions are the Goldstone modes corresponding to the
spontaneous chiral symmetry breaking and dominate the thermodynamics
of the system at low $T$ and low $\mu_B$. In the pion superfluid,
however, mesons $\sigma, \pi_+$ and $\pi_-$ are coupled together and
no longer the eigen modes of the Hamiltonian of the system, the new
eigen modes, labeled by $\overline\sigma, \overline\pi_+$ and
$\overline\pi_-$ are linear combinations of $\sigma, \pi_+$ and
$\pi_-$. The pole equation (\ref{pole1}) in this case is separated
into $1-2G\Pi_{\pi_0\pi_0}\left(M_{\overline{\pi}_0}^2, 0\right)=0$
for $\overline{\pi}_0$ (to avoid confusion with the $\pi_0$ in
normal phase, we still use $\overline\pi_0$ to stand for $\pi_0$ in
the pion superfluid) and
$\text{det}\left[1-2G\widetilde\Pi\left(M_j^2,0\right)\right]=0$ for
$\overline\sigma, \overline\pi_+$ and $\overline\pi_-$, where the
polarization $\widetilde\Pi$ is the sub-matrix of $\Pi$ defined in
the meson subspace $\{\sigma, \pi_+, \pi_-\}$. The $T$, $\mu_B$ and
$\mu_I$ dependence of $M_j$ is shown in Ref.~\cite{he}.

From the Yukawa potential between two nucleons, $V(r)\sim e^{-{\cal
M} r}/r$, its strength is governed by the mass ${\cal M}$ of the
exchanged boson. The long range force is suppressed or screened by
the massive boson (this is the reason why ${\cal M}$ is called the
screening mass). Its inverse is the screening length ${\cal
R}=1/{\cal M}$. At finite temperature, the screening mass is defined
as the pole of the boson propagator at $q_0^2=0, {\bf q}^2=-{\cal
M}^2$. In the NJL model at quark level, the meson screening mass
${\cal M}$ which control the quark potential are defined as
\begin{equation}
\label{pole2} \text{det}\left[1-2G\Pi\left(0,-{\cal
M}_j^2\right)\right]=0,
\end{equation}
which is again simplified to four independent equations
$1-2G\Pi_{jj}(0,-{\cal M}_j^2)=0$ in normal matter for $j=\sigma,
\pi_+, \pi_-, \pi_0$ and separated into
$1-2G\Pi_{\pi_0\pi_0}(0,-{\cal M}_{\overline\pi_0}^2)=0$ for
$\overline\pi_0$ and $\text{det}[1-2G\widetilde\Pi(0,-{\cal
M}_j^2)]=0$ in the meson subspace for $\overline\sigma,
\overline\pi_+$ and $\overline\pi_-$ in the pion superfluid.

Before the numerical calculation, we can analytically prove that in
the pion superfluid the mass equation (\ref{pole1}) at $M=0$ or
(\ref{pole2}) at ${\cal M}=0$ becomes exactly the second gap
equation of (\ref{gaps}) for the pion condensate,
\begin{equation}
\label{goldstone} \text{det}\left[1-2G\Pi\left(0,0\right)\right]=0.
\end{equation}
This means that the Goldstone mode with $M={\cal M}=0$,
corresponding to the spontaneous isospin symmetry breaking, is
automatically guaranteed in the RPA approximation in the NJL model,
and the Yukawa potential via the exchange of the Goldstone meson is
not screened.

There are three parameters in the NJL model, the current quark mass
$m_0$, the coupling constant $G$ and the momentum cutoff $\Lambda$.
In the following numerical calculations, we take $m_0$=5 MeV,
$G$=4.93 GeV$^{-2}$ and $\Lambda$=653 MeV. This group of parameters
corresponds to the pion mass $M_\pi$=134 MeV, the pion decay
constant $f_\pi$=93 MeV and the effective quark mass $m$=310 MeV in
vacuum. To extract reliable conclusions from the numerical
calculations in the model, the thermodynamic parameters, namely the
temperature and quark chemical potentials, should be much less than
the cutoff, $T, \mu_B/3, \mu_I/2 \ll \Lambda$. In our following
numerical calculations we take $T<300$ MeV, $\mu_B<900$ MeV and
$\mu_I<600$ MeV. Considering the critical isospin chemical potential
$\mu_I^c=134$ MeV~\cite{he}, $\mu_I=600$ MeV $\simeq 4 \mu_I^c$ is
large enough for the discussion of pion superfluid.

\begin{figure}[!hbt]
\includegraphics[width=0.55\textwidth]{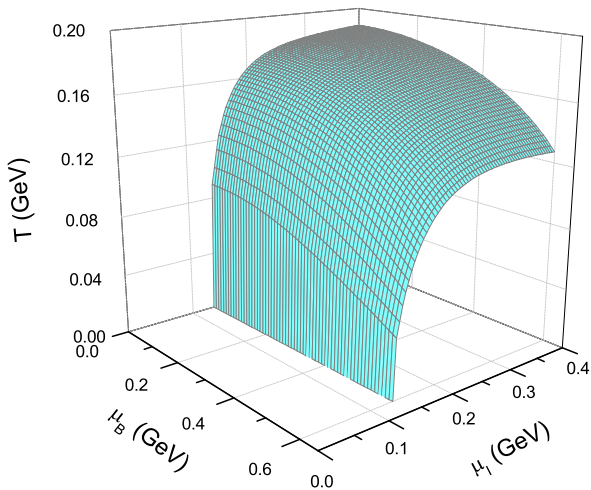}
\caption{The phase transition hypersurface of pion superfluid in the
three dimensional space of temperature $T$ and baryon and isospin
chemical potentials $\mu_B$ and $\mu_I$ in the NJL model. }
\label{fig1}
\end{figure}

With the coupled gap equations (\ref{gaps}) for the chiral and pion
condensates, the phase boundary of the pion superfluid as a
hypersurface in the three dimensional space of temperature and
baryon and isospin chemical potentials is shown in Fig.~\ref{fig1}.
Since the pion condensate increases with decreasing temperature and
baryon density and increasing isospin density, the pion superfluid
phase is located in the region of low temperature, low baryon
density and high isospin density. By comparing the second gap
equation of (\ref{gaps}) for the pion condensate with the dynamical
mass equation (\ref{pole1}) for pions, the critical isospin chemical
potential for the pion superfluid at $T=\mu_B=0$ is exactly the pion
mass in vacuum, $\mu_I^c=M_\pi=134$ MeV. This critical value
increases with increasing temperature and baryon chemical potential.
%%%%%%%%%%%%%%%%%%%%%%%%%%%%%%%%%%%%%%%%%%%%%%%%%%%%%%%%%%%%%%%%
%%%%%%%%%%%%%%%%%%%%%%%%%%%%%%%%%%%%%%%%%%%%%%%%%%%%%%%%%%%%%%%
\begin{figure}[!hbt]
\includegraphics[width=0.48\textwidth]{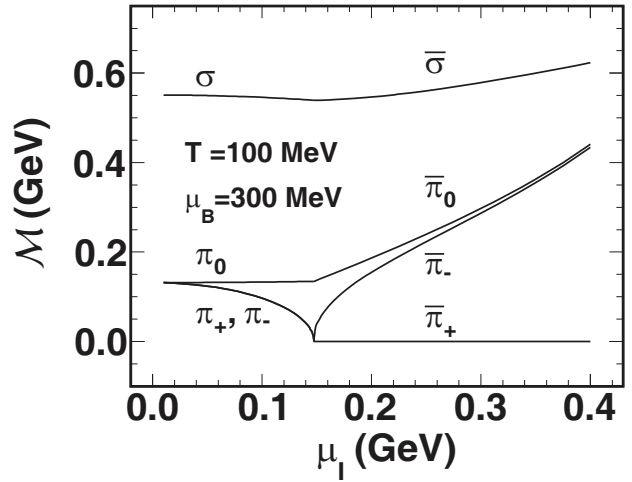}
\caption{The meson screening masses ${\cal M}$ as functions of
isospin chemical potential $\mu_I$ at temperature $T=100$ MeV and
baryon chemical potential $\mu_B=300$ MeV. } \label{fig2}
\end{figure}

The isospin chemical potential dependence of the screening masses at
fixed temperature and baryon chemical potential is shown in
Fig.~\ref{fig2}. For the chosen $T=100$ MeV and $\mu_B=300$ MeV, the
critical $\mu_I$ for the pion superfluid is $\mu_I^c=142$ MeV. Since
mesons carry isospin charge, the isospin dependence of the meson
propagator $D$ comes from not only the chiral and pion condensates
$\sigma(\mu_I)$ and $\pi(\mu_I)$ but also an explicit shift for the
meson energy $q_0\to q_0-\mu_I^j$, $D((q_0-\mu_I^j)^2,{\bf
q}^2;\sigma(\mu_I),\pi(\mu_I))$, where $\mu_I^j$ is the meson
isospin chemical potential. This explicit $\mu_I$ dependence leads
to a dynamical mass splitting for the charged pions~\cite{he} even
in the normal phase. For the screening mass defined at $q_0^2=0,
{\bf q}^2=-{\cal M}^2$ of the propagator, however, the explicit and
inexplicit isospin dependence in the normal phase is the same for
the two charged pions, and therefore they are degenerate in the
normal phase. Considering the condensation of $\pi_+$ at $\mu_I >
\mu_I^c$, ${\cal M}_{\overline\pi_+}$ and ${\cal
M}_{\overline\pi_-}$ are different from each other in the pion
superfluid. The degeneracy of ${\cal M}_{\pi_+}$ and ${\cal
M}_{\pi_-}$ in the normal phase leads to two zero modes at the
critical point, which is the reason why the strength of the quark
potential approaches to the maximum at the phase transition very
fast, see the discussion below. In the pion superfluid, only $\pi_0$
is still the eigen mode, but $\sigma, \pi_+$ and $\pi_-$ are
replaced by the new eigen modes $\overline \sigma, \overline \pi_+$
and $\overline \pi_-$. $\overline \pi_+$ is the Goldstone mode with
zero screening mass ${\cal M}_{\pi_+}=0$, as we analyzed above. At
small isospin chemical potential $\mu_I<\mu_I^c$, the chiral
symmetry $SU_A$(2) is already explicitly broken to $U_A$(1).
Therefore, the chiral symmetry restoration at high isospin chemical
potential means only degeneracy of $\sigma$ and $\pi_0$, the charged
$\pi_+$ and $\pi_-$ behave differently. However, in Fig.~\ref{fig2}
the degeneracy at high isospin chemical potential is not for
$\overline\pi_0$ and $\overline\sigma$ but for $\overline\pi_0$ and
$\overline\pi_-$. This is due to the strong mixing among $\sigma,
\pi_+$ and $\pi_-$~\cite{he}.
%%%%%%%%%%%%%%%%%%%%%%%%%%%%%%%%%%%%%%%%%%%%%%%%%%%%%%%%%%%%%%%
\begin{figure}[!hbt]
\includegraphics[width=0.52\textwidth]{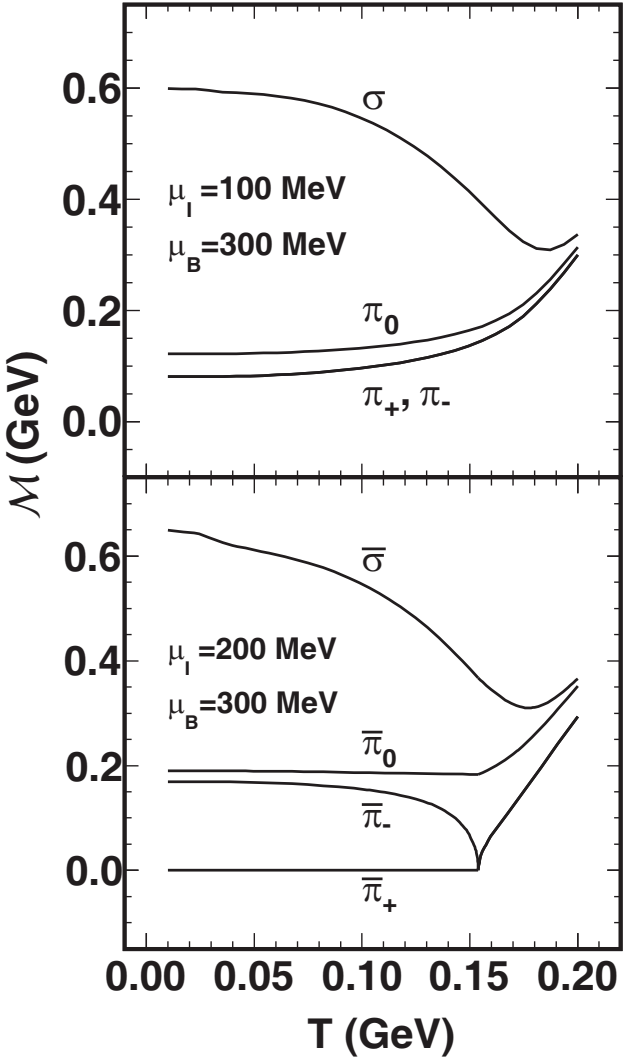}
\caption{The meson screening masses ${\cal M}$ as functions of
temperature $T$ in normal phase (top panel) at isospin chemical
potential $\mu_I=100$ MeV and in pion superfluid (bottom panel) at
$\mu_I=200$ MeV. In both phases the baryon chemical potential is
fixed to be $\mu_B=300$ MeV. } \label{fig3}
\end{figure}

In the top panel of Fig.~\ref{fig3} we show the screening masses in
the normal phase as functions of temperature at fixed baryon and
isospin chemical potentials $\mu_B=300$ MeV and $\mu_I=100$ MeV. The
charged pions $\pi_+$ and $\pi_-$ coincide with each other in the
whole plane, as we discussed above. Since there is no $\sigma - \pi$
mixing in the normal phase, the $\sigma$ and $\pi_0$ mesons become
degenerate at high temperature where the chiral symmetry is almost
restored. For a small isospin chemical potential, the mass splitting
between neutral and charged pions is slight.

In order to clearly see the effect of spontaneous and explicit
isospin symmetry breaking on the screening masses, we show in the
bottom panel of Fig.~\ref{fig3} their temperature dependence at a
higher isospin chemical potential. For the chosen $\mu_I=200$ MeV
and $\mu_B=300$ MeV, the critical temperature for the phase
transition from pion superfluid to normal phase is $T_c\simeq 155$
MeV. In the pion superfluid at $T<T_c$, $\overline\pi_+$ is the
Goldstone boson with screening mass ${\cal M}_{\overline\pi_+}=0$.
Because of the mixing among $\sigma, \pi_+$ and $\pi_-$,
$\overline\pi_0$ and $\overline\pi_-$, instead of $\sigma$ and
$\pi_0$, approach to each other at low temperature where the mixing
is strong enough. When $T$ is close to the critical point,
$\overline\pi_-$ approaches to the Goldstone mode due to the
weakening of the pion condensation. In normal phase at $T>T_c$,
$\pi_+$ and $\pi_-$ coincide again, and with increasing temperature
$\sigma$ and $\pi_0$ mesons approach to each other when the chiral
symmetry is gradually restored.

\section{quark potential}
\label{s3} By analogy with the Yukawa potential between two nucleons
through one boson exchange, the static quark potential via one meson
exchange can be expressed as the Fourier transform of the meson
propagator at $q_0=0$,
\begin{equation}
\label{v1} V(r)=\int{d^3 {\bf q}\over (2\pi)^3}D(0,{\bf
q}^2)e^{i{\bf q}\cdot{\bf r}}.
\end{equation}
In the flavor SU(2) NJL model, the meson propagator in the pion
superfluid is a $4\times 4$ matrix with off diagonal elements, the
quark potential becomes
\begin{eqnarray}
\label{v2} V(r)&=&\int{d^3 {\bf q}\over (2\pi)^3}\text {Tr} D(0,{\bf
q}^2)e^{i{\bf q}\cdot{\bf r}}\nonumber\\
&=&-{G\over \pi^2 r}\int dq\text{Tr}{q\sin (qr)\over
1-2G\Pi(0,q^2)},
\end{eqnarray}
where the trace is taken in the meson space
$\{\sigma,\pi_+,\pi_-,\pi_0\}$. Note that there is no singularity
for the propagator on the real momentum axis and the above
integration is numerically straightforward without any technical
problem.

To better understand the potential through the screening of the
interaction, we take also the pole approximation in calculating the
quark potential. In this case the potential can be simplified as a
summation over the poles,
\begin{eqnarray}
\label{v3} V(r) &=& -\int {d^3{\bf q}\over (2\pi)^3}\sum_j{g_{jq\bar
q}^2\over q^2+{\cal M}_j^2}e^{i{\bf q}\cdot{\bf r}}\nonumber\\
&=&-\sum_j{g_{jq\bar q}^2\over 4\pi r}e^{-{\cal M}_j r}
\end{eqnarray}
with $j=\sigma,\pi_+,\pi_-,\pi_0$ in the normal phase and
$j=\overline\sigma,\overline\pi_+,\overline\pi_-,\overline\pi_0$ in
the pion superfluid, where $g_{jq\bar q}$ is the effective
meson-quark-antiquark coupling constant,
\begin{equation}
\label{coupling1} g_{jq\bar q}^2 = {2G\sum_i\left[{\text
{det}\left[1-2G\Pi(0,q^2)\right]\over
1-2G\Pi(0,q^2)}\right]_{ii}\over {d\over d
q^2}\text{det}\left[1-2G\Pi(0,q^2)\right]}\Bigg|_{q^2=-{\cal M}_j^2}
\end{equation}
which is reduced to
\begin{equation}
\label{coupling2} g_{jq\bar q}^2 = -\left[{d\Pi_{jj}(0,q^2)\over
dq^2}\right]^{-1}_{q^2=-{\cal M}_j^2}
\end{equation}
when the pion condensation disappears.
%%%%%%%%%%%%%%%%%%%%%%%%%%%%%%%%%%%%%%%%%%%%%%%%%%%%%%%%%%%%%%%%%%
\begin{figure}[!hbt]
\includegraphics[width=0.48\textwidth]{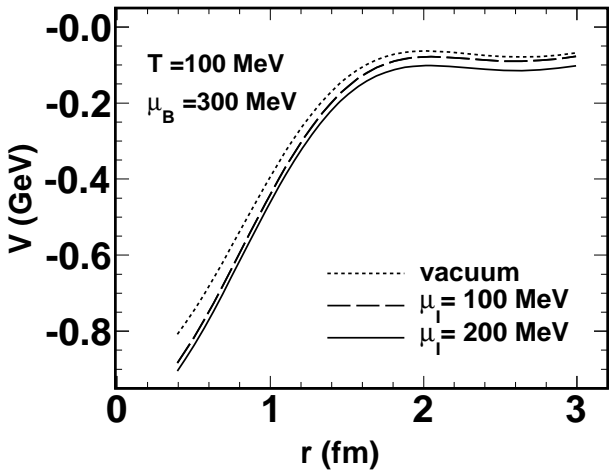}
\caption{The quark potential as a function of the distance $r$
between the two quarks in normal phase at $\mu_I=100$ MeV and in
pion superfluid at $\mu_I=200$ MeV. In both phases the temperature
and baryon chemical potential are fixed to be $T=100$ MeV and
$\mu_B=300$ MeV. The dotted line is the potential in vacuum.}
\label{fig4}
\end{figure}
%%%%%%%%%%%%%%%%%%%%%%%%%%%%%%%%%%%%%%%%%%%%%%%%%%%%%%%%%%%%%%%%
\begin{figure}[!hbt]
\includegraphics[width=0.48\textwidth]{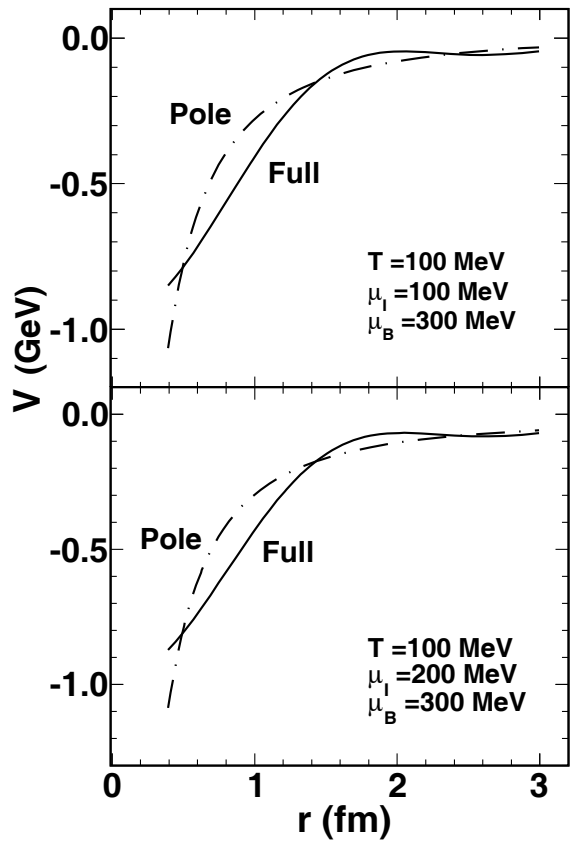}
\caption{The quark potential calculated exactly and in pole
approximation in normal matter (top panel) with $\mu_I=100$ MeV and
in pion superfluid (bottom panel) with $\mu_I=200$ MeV. In both
phases the temperature and baryon chemical potential are fixed to be
$T=100$ MeV and $\mu_B=300$ MeV. } \label{fig5}
\end{figure}

In the normal phase at zero isospin density, it is easy to expect
that the quark potential is gradually weaken by finite temperature
and baryon density, and this is confirmed by the lattice
simulation~\cite{kaczmarek} and model calculation~\cite{mu2}. What
we focus on in this paper is the isospin effect on the quark
potential. In Fig.~\ref{fig4} we show the potential as a function of
the distance $r$ between the two quarks in normal phase with
$\mu_I=100$ MeV (dashed line) and in pion superfluid with
$\mu_I=200$ MeV (solid line), the temperature and baryon chemical
potential are fixed to be $T=100$ MeV and $\mu_B=300$ MeV in the two
phases. Considering the momentum cutoff in the model, we do not
discuss the potential in short distance and take the starting point
$r=0.4$ fm. The effect of the finite cutoff on the thermodynamics of
the system is mainly at high temperature\cite{zhuang1}. While this
effect is not remarkable in the temperature and density region we
considered here, the potential is normalized to
$\lim_{T\to\infty}V(r)=0$ in the high temperature limit. It is easy
to see that this normalization is automatically assumed in the pole
approximation (\ref{v3}). In both the normal phase and pion
superfluid, the potentials oscillate slightly at large distance.
This is from the so called Friedel oscillation~\cite{kohn} induced
by the sharp Fermi surface at zero temperature which has been widely
discussed in different matters~\cite{kapusta,alonso,shuryak}. Since
the Fermi surface is smeared at finite temperature, the oscillation
is suppressed by temperature and will disappear when the temperature
is high enough. The difference between the normal matter and pion
superfluid can be understood clearly in the pole approximation
(\ref{v3}). From the screening masses shown in Figs.~\ref{fig2} and
\ref{fig3}, $\sigma\ (\overline\sigma)$ is always much heavier than
the other mesons in the region we discussed, its contribution to the
potential is almost screened, and in the intermediate and large
distance the potential is mainly from the pion exchange. From
Fig.\ref{fig2}, the pion screened masses at $\mu_I=100$ and $200$
MeV satisfy the relations ${\cal M}_{\pi_0} < {\cal
M}_{\overline\pi_0}$ and ${\cal M}_{\pi_-} < {\cal
M}_{\overline\pi_-}$. If we neglect $\pi_+$ and $\overline \pi_+$,
the potential in the pion superfluid should be above the potential
in normal matter. However, when $\pi_+$ and $\overline\pi_+$ are
included, the mass relation ${\cal M}_{\pi_+} > {\cal
M}_{\overline\pi_+}=0$ may turn around the above isospin dependence,
and the attractive potential in the pion superfluid may be stronger
than the one in normal matter, namely the potential with $\mu_I=200$
MeV may be below the potential with $\mu_I=100$ MeV. This is
confirmed from our numerical calculation, as shown in
Fig.~\ref{fig4}. While for $\mu_I=100$ and $200$ MeV the difference
between the two potentials is small, the trend of the isospin
dependence is really controlled by the Goldstone mode. We plotted
also the potential in vacuum (dotted line) in Fig.\ref{fig4}.
Totally different from the temperature and baryon density effect
which weakens the interaction among quarks, the potential strength
increases with isospin chemical potential.
%{\color{red} Since
%the medium effect is a long distance effect, its influence on the
%quark potential can not be seen in short distance. This can be
%understood from the pole approximation (\ref{v3}), the temperature
%and density effect which is fully reflected in the screening masses
%in the exponential function plays important role only for large $r$,
%as clearly shown in Fig.\ref{fig4} where the two curves with
%different medium effects coincide for small $r$.}

The above numerical calculation for the potential is full but the
analysis on its isospin dependence is based on the behavior of the
screening masses. To know to what extent this analysis is valid, we
compare in Fig.~\ref{fig5} the exact calculation with the pole
approximation in normal matter with $\mu_I=100$ MeV and in pion
superfluid with $\mu_I=200$ MeV. Again the temperature and baryon
chemical potential are fixed to be $T=100$ MeV and $\mu_B=300$ MeV
in the two phases. While the pole approximation deviates from the
full result remarkably in the region of $r<2$ fm, and the Friedel
oscillation is washed away in the approximation, the screening
masses do qualitatively describe the potential, for instance, the
exponential decay and the saturation at large distance.

\begin{figure}[!hbt]
\includegraphics[width=0.5\textwidth]{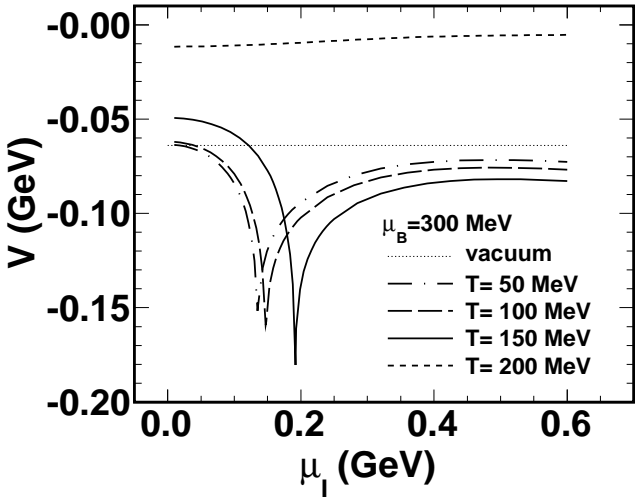}
\caption{The quark potential at fixed distance $r=2$ fm as a
function of isospin chemical potential $\mu_I$ at fixed baryon
chemical potential $\mu_B=300$ MeV and for different temperature
$T$. } \label{fig6}
\end{figure}

To see clearly and comprehensively the medium effect, we show in
Fig.~\ref{fig6} the isospin dependence of the quark potential at a
fixed distance. We take $r=2$ fm which is the typical diameter of a
hadron, the potential at this distance can be used to describe the
average medium effect on the coupling strength of the quark matter.
From the phase diagram Fig.~\ref{fig1}, for the chosen baryon
chemical potential $\mu_B=300$ MeV and the isospin region
$\mu_I<600$ MeV, the maximum critical temperature for the pion
superfluid is $T_c\simeq 180$ MeV.
%%%%%%%%%%%%%%%%%%%%%%%%%%%%%%%%%%%%%%%%%%%%%%%%%%%%%%%%%%%%%%%%
%%%%%%%%%%%%%%%%%%%%%%%%%%%%%%%%%%%%%%%%%%%%%%%%%%%%%%%%%%%%%%%%
\begin{figure}[!hbt]
\includegraphics[width=0.5\textwidth]{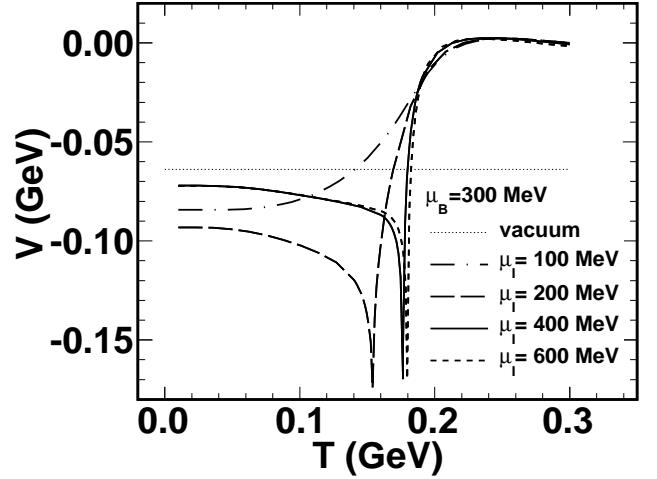}
\caption{The quark potential at fixed distance $r=2$ fm as a
function of temperature $T$ at fixed baryon chemical potential
$\mu_B=300$ MeV and for different isospin chemical potential
$\mu_I$. } \label{fig7}
\end{figure}

Let us first look at a potential at a given temperature. For any
$T<T_c$, the strength of the attractive potential increases with
increasing $\mu_I$ in normal phase and approaches to the maximum at
the critical value $\mu_I^c$. When the system enters the pion
superfluid, the behavior of the potential suddenly changes, its
strength turns from increasing to decreasing and gets saturated
fast. The sharp and deep valley at the critical point $\mu_I^c$
arises from the two zero modes $\pi_+$ and $\pi_-$ or $\overline
\pi_+$ and $\overline\pi_-$ at the phase transition, shown in
Figs.~\ref{fig2} and \ref{fig3}. Note that the saturated potential
in the pion superfluid is always below the vacuum potential $V(r=2\
\text{fm})\simeq -0.06$ GeV shown as dotted line in Fig.\ref{fig6}.
Therefore, from the valley structure of the potential, the quark
matter is most strongly coupled at the phase transition point, and
from the strong enough potential at high isospin density, the pion
superfluid is always strongly coupled. At $T>T_c$, for instance
$T=200$ MeV, there is no more phase transition, and the valley
structure disappears. At such high temperature, the potential at the
distance of $r=2$ fm almost disappears.

We now focus on the temperature dependence of the potential at a
given isospin chemical potential, shown in Fig.\ref{fig7}. For a
small $\mu_I<134$ MeV, there is no pion superfluid even at zero
temperature, the system is always in the normal phase. In this case,
the temperature effect is under our expectation, the potential
becomes weaker and weaker in the hot medium and finally vanishes.
For a large $\mu_I>134$ MeV, the system is in pion superfluid at
$T<T_c$ and normal phase at $T>T_c$. At a fixed isospin chemical
potential, there is also a valley structure around the critical
temperature: The potential strength increases with increasing
temperature and reaches the maximum at $T_c$, and after crossing the
phase transition point, the potential changes suddenly and turns to
become more and more weak and finally disappears. Different from the
isospin structure shown in Fig.\ref{fig6}, the potential goes to
zero very fast in the normal phase.

\section{Conclusion}
\label{s4} We have investigated the meson screening masses and quark
potential in a pion superfluid in the frame of NJL model. The
Goldstone mode corresponding to the spontaneous isospin symmetry
breaking plays a significant role in the thermodynamics of the
system. The minimum of the attractive quark potential is always
located on the phase transition hypersurface in the three
dimensional space of temperature and baryon and isospin chemical
potentials. While at extremely high temperature and baryon density,
quark matter is expected to be in a weakly coupled state, the pion
superfluid is in a strongly coupled phase even the isospin density
is extremely high. These surprising properties of pion superfluid
may help us to understand the isospin asymmetric matter, like the
core of compact stars.

In this paper, we considered quarks in mean field approximation and
mesons in RPA. When we go beyond the mean field for the pion
condensate by taking into account the feedback from the mesons, the
phase boundary will be shifted and the quark potential in the region
of low temperature and low densities will be modified, like the case
of chiral phase transition~\cite{zhuang2}. However, the minimum of
the potential at the phase transition and the non-zero potential at
extremely high isospin density will remain, since they are
controlled by the Goldstone mode.

\appendix {\bf Acknowledgement:} The work is supported by the NSFC
grant Nos. 10735040, 10847001, 10975084 and 11079024.

\end{document}